\documentclass{PoS}

\usepackage{amssymb,amsfonts,amsmath,amsthm}   
\usepackage{dsfont}                            

\title{Moments of nucleon distribution amplitudes from irreducible three-quark operators}

\ShortTitle{Moments of nucleon distribution amplitudes}

\author{M.~G\"ockeler$^a$, R.~Horsley$^b$, \speaker{T.~Kaltenbrunner\footnote{Speaker (Irreducible three quark operators for LQCD)}}~$^a$, Y.~Nakamura$^c$, D.~Pleiter$^c$, P.~E.~L.~Rakow$^d$, A.~Sch\"afer$^a$, G.~Schierholz$^{c,e}$, H.~St\"uben$^f$, \speaker{ N.~Warkentin\footnote{Speaker (Moments of leading-twist and next-to-leading twist nucleon distribution amplitudes)}}~$^a$ and J.~M.~Zanotti$^b$\\%
        \llap{$^a$} Institut f\"ur Theoretische Physik, Universit\"at Regensburg, 93040 Regensburg, Germany\\
        \llap{$^b$} School of Physics, University of Edinburgh, Edinburgh EH9~3JZ, UK\\
        \llap{$^c$} John von Neumann Institute NIC/DESY Zeuthen, 15738 Zeuthen, Germany\\
        \llap{$^d$} Theoretical Physics Division, Department of Mathematical Sciences, University of Liverpool, Liverpool L69~3BX, UK\\
        \llap{$^e$} Deutsches Elektronen-Synchrotron DESY, 22603 Hamburg, Germany\\
        \llap{$^f$} Konrad-Zuse-Zentrum f\"ur Informationstechnik Berlin, 14195 Berlin, Germany\\
        E-mail: \email{Thomas.Kaltenbrunner@physik.uni-regensburg.de}\\ 
        E-mail: \email{Nikolaus.Warkentin@physik.uni-regensburg.de} \\
      }

\author{QCDSF collaboration}
\abstract{
Semi-exclusive and exclusive processes are becoming more and more important in high energy physics since they are excellently suited to study the internal hadronic structure. To analyze such processes the knowledge of the hadron distribution amplitudes, which are universal for different reactions, is essential. Only rather indirect information on these nonperturbative functions can be obtained from measurements. In this work we report on a lattice QCD computation of moments of nucleon distribution amplitudes using suitable three-quark operators. 
However, these operators have to be renormalized and the mixing is even more complicated than in the continuum. Using the symmetry group of the hypercubic lattice we therefore derive and implement irreducibly transforming three-quark operators, which allow us to control the mixing pattern and will finally lead to quantitative predictions in the $\overline{\mathrm{MS}}$ scheme. We present preliminary results for leading-twist and next-to-leading twist nucleon distribution amplitudes based on the QCDSF/UKQCD simulations with 2 flavors of dynamical clover fermions.
}

\dedicated{Edinburg 2007/33\\
Liverpool LTH~765}

\FullConference{The XXV International Symposium on Lattice Field Theory\\
		 July 30 - August 4 2007\\
		 Regensburg, Germany}

\begin{document}

\section{Introduction}

QCD is the theory of the strong interaction, but nevertheless many results are obtained from perturbative calculations applicable to a large variety of scattering processes. The success of this approach is based on the factorization properties of the investigated reactions, which allow one to introduce distribution and fragmentation functions for quarks and gluons in the case of inclusive, and quark distribution amplitudes in the case of hard exclusive processes. These universal functions are common to different processes in all orders of the perturbative expansion. In ongoing and future experiments the investigation of hard exclusive and semi-exclusive processes will become the key tool for increasing our knowledge and understanding of the internal and spin structure of hadrons.
Thus for facilitating fully quantitative predictions for these processes the knowledge of the nonperturbative quark distribution amplitudes is essential. They describe the hadron structure in terms of valence quark Fock states at small transverse separation and, unlike distribution functions in inclusive processes, cannot be accessed directly in experiment. Only some indirect insight can be obtained by measuring physical quantities like the magnetic form factor of the nucleon $G_M(Q^2)$. At very large values of $Q^2$ the electromagnetic form factors of the nucleon can be expressed as a convolution of the hard scattering kernel $h(x_i, y_i, Q^2)$ and the quark distribution amplitude in the nucleon $\varphi(x_i, Q^2)$ \cite{Lepage:1979za}:
   \begin{equation}
      G_M(Q^2)=\int_0^1 [\mathrm dx] \int_0^1 [\mathrm dy]
               \varphi^\star(y_i,Q^2) h(x_i, y_i, Q^2) \varphi(x_i,Q^2)
               +O(m^2/Q^2),
   \end{equation}  
where $[\mathrm dx]=\mathrm dx_1\mathrm dx_2 \mathrm dx_3\delta(1-\sum_{i=1}^3 x_i)$, and $Q^2$ equals the modulus of the squared momentum transfer in the hard scattering process. In this case only the leading twist nucleon distribution amplitude contributes. In an appropriate gauge,  $x_i$ ($y_i$) can be interpreted as the momentum fractions carried by the valence quarks before (after) the hard scattering.

Apart from QCD sum rule determinations, an analytic approach to the distribution amplitude is feasible only for sufficiently large values of $Q^2$, where the asymptotic form $\varphi(Q^2\rightarrow\infty)=120x_1x_2x_3$ \cite{Lepage:1980fj, Lepage:1979zb} is obtained. However, given the logarithmic evolution in $Q^2$ this knowledge is not really useful at reasonable energy scales, such that a nonperturbative lattice calculation seems to be the method of choice. 

At intermediate values of the momentum transfer ($1 \leq Q^2 \leq 10\, \textrm{GeV}^2$) the electromagnetic form factors can be calculated from the nucleon distribution amplitudes using lightcone sum rules. In this case also higher twist terms of the nucleon distribution amplitudes \cite{Braun:2000kw,Braun:2001tj,Huang:2004vf,Braun:2006hz}  will contribute. This kinematic region  gained a lot of interest in recent years, because new data from JLAB \cite{Jones:1999rz,Gayou:2001qt,Gayou:2001qd,Punjabi:2005wq}   for the well-known electromagnetic form factors of the nucleon contradict common textbook knowledge, for details see \cite{Perdrisat:2006hj} and references therein.

In this paper we want to present the theoretical framework needed to set up a calculation of the nucleon distribution amplitudes on the lattice and present some preliminary results for the leading and some next-to-leading twist distribution amplitudes.

\section{Theoretical background}
   \begin{figure}
   \centerline{
      \includegraphics[width=0.65\textwidth]{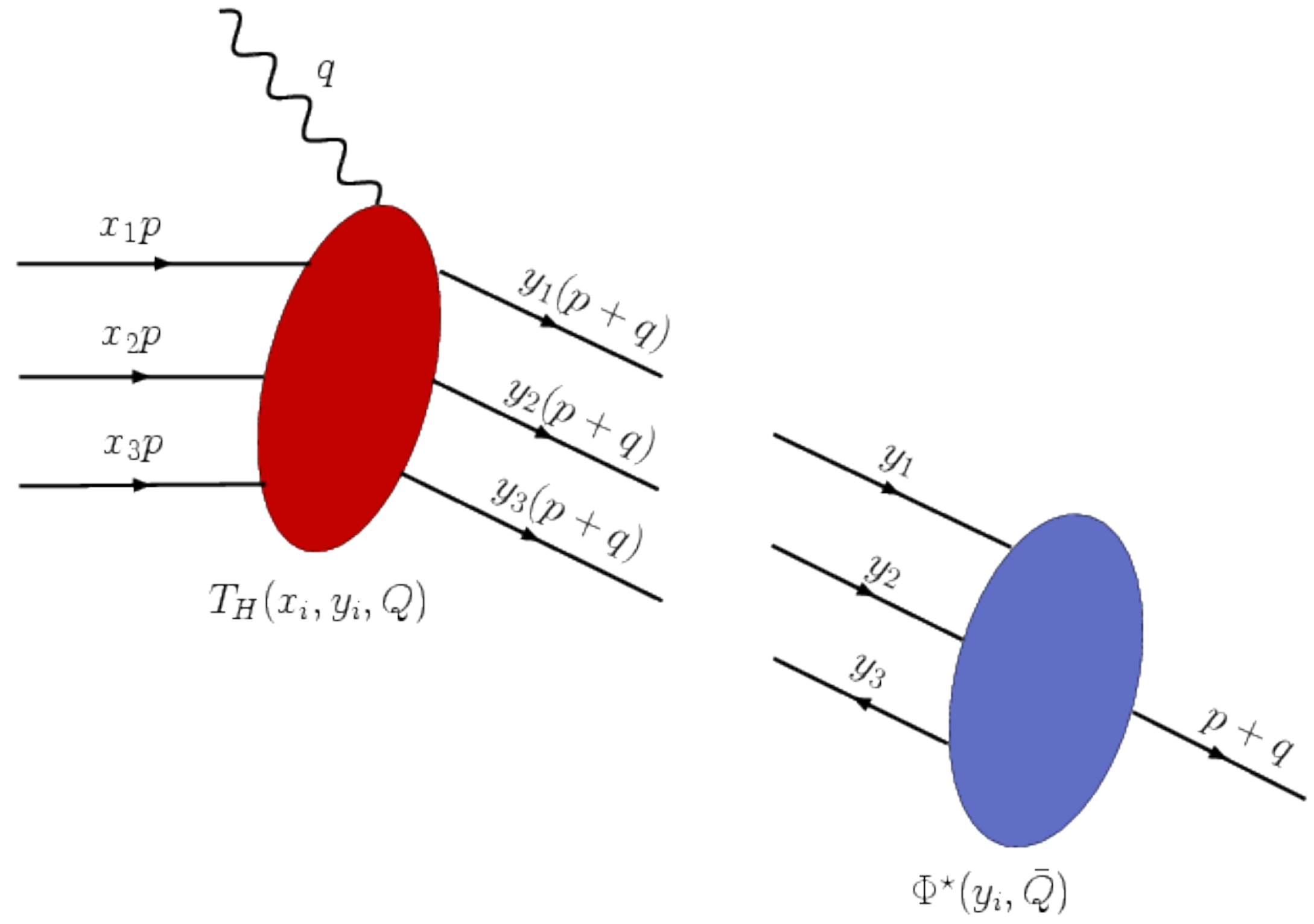}
   }
   \caption{\label{fig_fact} The factorization of the subprocess relevant in the calculations involving the quark distribution amplitudes of the nucleon}
   \end{figure}
We would like to stress once more that the nonperturbative nucleon distribution amplitudes are needed anytime a process is calculated in which three quarks form a nucleon, e.g., after a hard scattering process as displayed in Fig.~\ref{fig_fact}. In Minkowski space
our starting point for the derivation of quark distribution amplitudes is the matrix element of a tri-local operator,
   \begin{equation}
   \label{eq_trilocal}
      \langle 0 \vert
      \left[
         \exp \left( ig \int_{z_1}^{z_3} A_\mu(\sigma) \mathrm d \sigma^\mu  \right)
         u_\alpha (z_1)
      \right]^a
      \left[
         \exp \left( ig \int_{z_2}^{z_3} A_\nu(\tau) \mathrm d \tau^\nu  \right)
         u_\beta(z_2)
      \right]^b \times
      d_\gamma^c(z_3)\;
      \vert N(p) \rangle \epsilon^{abc},
   \end{equation} 
where path ordering is implied for the exponentials, $a,b,c$ are the color indices and $\vert N(p)\rangle$ denotes the nucleon state with momentum $p$. We will consider these matrix elements for space-time separations of the quarks on the light cone with $z_i=a_i z$ ($z^2=0$) and $\sum_i a_i=1$.

Using the transformation properties of the fields in eq.~(\ref{eq_trilocal}) under Lorentz symmetry and parity it is possible to rewrite the leading twist contribution in terms of three invariant functions $V$, $A$ and $T$ \cite{Henriques:1975uh},
   \begin{equation}
      \begin{split}
      (\text{\ref{eq_trilocal}})=
      \frac{1}{4} f_N 
      \left\{
            (p\cdot\gamma C)_{\alpha\beta} (\gamma_5 N)_\gamma V(z_i\cdot p)+ (p\cdot \gamma \gamma_5 C)_{\alpha\beta} N_\gamma A(z_i\cdot p)
      \right.\\
      \left.
            +(i\sigma_{\mu\nu}p^\nu C)_{\alpha\beta}(\gamma^\mu\gamma_5 N)_\gamma T(z_i\cdot p)
      \right\},
      \end{split}
      \label{eq_dadef}
   \end{equation} 
where $C$ is the charge conjugation matrix, $N$ the nucleon spinor and $f_N$ the nucleon decay constant. Beyond the large $Q^2$ limit there are also higher twist contributions, which we do not discuss here.

In momentum space 
   \begin{equation}
      V(x_i)=\int V(z_i\cdot p) \prod_{i=1}^{3}  \exp\left(ix_i (z_i\cdot p)\right) \frac{\mathrm d(z_i\cdot p)}{2\pi},
      \qquad V(x_i)\equiv V(x_1,x_2,x_3)
      \label{eq_daft}
   \end{equation} 
the distribution amplitudes $V(x_i)$, $A(x_i)$ and $T(x_i)$ describe the quark distribution inside the nucleon as a function of the longitudinal momentum fractions $x_i$. To be precise we want to note that the distribution amplitudes also depend on the factorization and renormalization scales. Here it is sufficient to set both scales to the same value $\mu$.

Since so far it is not possible to access the quark distribution amplitudes directly on the lattice we consider their moments, which are defined as
   \begin{equation}
      V^{lmn}=\int_0^1 [\mathrm dx]\; x_1^l x_2^m x_3^n\; V(x_1, x_2, x_3)
   \end{equation} 
with equivalent definitions for the other distribution amplitudes. Using eq.~(\ref{eq_dadef}) and (\ref{eq_daft}) one can relate these moments of the quark distribution amplitudes to matrix elements of the following local operators
   \begin{align}
      \mathcal V_\tau^{\rho \bar l \bar m \bar n }(0)=\epsilon^{abc} &
                                                                  [i^l \mathcal D^{\lambda_1}\dots \mathcal D^{\lambda_l}u^a_\alpha(0)] 
                                                                  (C\gamma^\rho)_{\alpha\beta}
                                                                  [i^m \mathcal D^{\mu_1}\dots \mathcal D^{\mu_m} u^b_\beta(0)] \nonumber\\
                                                                  & \times [i^n \mathcal D^{\nu_1}\dots \mathcal D^{\nu_n} (\gamma_5 d^c(0) )_{\tau} ]\label{eq_dav}\\
      \mathcal A_\tau^{\rho \bar l \bar m \bar n}(0)=\epsilon^{abc}&
                                                                  [(i^l \mathcal D^{\lambda_1}\dots \mathcal D^{\lambda_l} u_\alpha^a(0)]
                                                                     (C\gamma^\rho\gamma_5)_{\alpha\beta}
                                                                  [i^m \mathcal D^{\mu_1}\dots \mathcal D^{\mu_m} u_\beta^b(0)]\nonumber\\
                                                                  &\times[i^n \mathcal D^{\nu_1}\dots \mathcal D^{\nu_n}  d^c_\tau (0)]\label{eq_daa}\\
      \mathcal T_\tau^{\rho \bar l \bar m \bar n}(0)=\epsilon^{abc}&
                                                                  [i^l \mathcal D^{\lambda_1}\dots \mathcal D^{\lambda_l} u_\alpha^a(0)]
                                                                  \left(C(-i\sigma^{\xi\rho})\right)_{\alpha\beta}
                                                                  [i^m \mathcal D^{\mu_1}\dots \mathcal D^{\mu_m} u_\beta^b(0)]\nonumber\\
                                                                 & \times[i^n \mathcal D^{\nu_1}\dots \mathcal D^{\nu_n} (\gamma_\xi\gamma_5 d^c(0))_{\tau} ]\label{eq_dat}
  \end{align}
by
   \begin{align}
      \label{eq_opmatrix1}
      \langle 0\vert \mathcal V_\tau^{\rho \bar l \bar m \bar n }(0) \vert N(p)\rangle &=
         -f_N V^{lmn} p^\rho p^{\bar l} p^{\bar m} p^{\bar n}  N_\tau (p), \\
      \langle 0\vert \mathcal A_\tau^{\rho \bar l \bar m \bar n }(0) \vert N(p)\rangle &=
         -f_N A^{lmn} p^\rho p^{\bar l} p^{\bar m} p^{\bar n} N_\tau (p),\\
      \langle 0\vert \mathcal T_\tau^{\rho \bar l \bar m \bar n }(0) \vert N(p)\rangle &=
         2 f_N T^{lmn} p^\rho p^{\bar l} p^{\bar m} p^{\bar n} N_\tau (p).
      \label{eq_opmatrix3}
   \end{align}
The multiindex $ \bar l \bar m \bar n $ denotes the Lorentz structure given by the covariant derivatives $\mathcal D_{\mu} =\partial_\mu -igA_\mu$ in the local operator while $\rho$ and $\tau$ belong to a gamma matrix and a spinor respectively. The nucleon decay constant $f_N$ is normalized here by the choice $V^{000}=1$.

For a lattice calculation the above equations and operators are converted from Minkowski to Euclidean space-time, then the techniques known from meson and hadron spectroscopy can be used to determine the matrix elements. However in the discretized space-time we are faced with additional obstacles being absent in continuum calculations. Though not mentioned explicitly, the above equations contain renormalization factors and the results obtained on the lattice should be converted to an appropriate continuum renormalization scheme such as $\overline{\mathrm{MS}}$.  Due to the reduced symmetry of the discretized space-time we will have to treat additional mixings of operators with the same mass dimension which are forbidden in the continuum. Even worse, the discretized theory furthermore suffers from potential mixing with lower-dimensional operators. So it is important to notice that there is some freedom to choose especially well-suited operators exploiting the fact that different operators with distinct multiindices on the left hand side of eqs.~(\ref{eq_opmatrix1})-(\ref{eq_opmatrix3}) are related to the same moments on the right hand side.
Hence the first task is to find appropriate combinations of three-quark operators, for which the unwanted mixing is absent or at least strongly reduced.

\section{Irreducible representations of three-quark operators}
The matrix elements of the type $\langle 0 \vert \mathcal{O} \vert N \rangle$ (\ref{eq_opmatrix1})-(\ref{eq_opmatrix3}) considered in the previous section refer to quark distribution amplitudes for the nucleon. However we want to stress that the results derived in this section are more general and can be used in any kind of calculation involving local three-quark operators of the form
\begin{equation}
\mathcal{O}^{(i)}(x) = T^{(i)}_{\alpha \beta \gamma \mu_1 \dots \mu_n} u_\alpha(x) u_\beta(x) \mathcal{D}_{\mu_1} \dots \mathcal{D}_{\mu_n} d_\gamma(x),
\label{3qoperator}
\end{equation}
where $T^{(i)}$ is a tensor that represents the appropriate coefficients. As the actual position of the covariant derivatives does not influence the following discussion, we will assume that they act on the last quark unless stated otherwise. Isospin symmetrization will be discussed in detail in the following section.

These three-quark operators are subject to renormalization and possible mixing under renormalization. In order to get quantitative results from lattice simulations it is hence essential to perform a detailed study of their renormalization and mixing coefficients, preferably in a nonperturbative approach.
Let $\mathcal{O}^\text{bare}$ denote the lattice regularized, bare three-quark operator and $\mathcal{O}^{\text{ren}}$ its renormalized counterpart. They are related by a renormalization matrix $Z$:
   \begin{equation}
      \mathcal{O}^{(i),\text{ren}} = Z_{ij} \mathcal{O}^{(j),\text{bare}}.
   \end{equation}
Operator mixing shows up in non-vanishing off-diagonal elements of $Z$. Typically several hundred independent operators $\mathcal{O}^{(j)}$ may appear on the right hand side so that an elaborate approach is needed to gain control of the mixing issue. As in the case of quark-antiquark operators \cite{Gockeler:1996mu} the symmetry group $H(4)$ of the hypercubic lattice provides appropriate tools to reduce the dimension of the problem: One decomposes the operator space into subspaces transforming irreducibly with respect to the hypercubic group. Mixing under renormalization is then possible exclusively between equivalent irreducible representations, i.e., between operators that obey exactly the same transformation laws under the group action. Thus the $Z$-matrix becomes block diagonal and one has to care only about the lower-dimensional non-identical blocks, which typically mix an order of one to ten irreducible operators.

We employ a similar group-theoretical approach for our three-quark operators. However, since half integer spin is assigned to our operators, we have to use the double cover of $H(4)$, the so-called spinorial hypercubic group $\overline{H}(4)$ which was studied first by Dai and Song in 2001 \cite{Dai:2001zu}. This finite group contains 768 elements and can be defined by six generators, $t$, $\gamma$, $I_1$, $I_2$, $I_3$ and $I_4$, and a set of generating equations:
 \begin{align}
   I_i^2&=-1,           & I_iI_j&=-I_jI_i          &        tI_1&=I_1t,         & & &\nonumber\\
   tI_2&=I_4t,          & tI_3&=I_2t,              &        tI_4&=I_3t,         & \nonumber\\
   \gamma I_1 &= -I_3,  & \gamma I_2&= -I_2\gamma, & \gamma I_4 &= -I_4\gamma,  & & & \nonumber\\
   \gamma^2&=-1,        & t^3&=-1,                 & (t\gamma)^4&=-1.           & & &\nonumber
 \end{align}
Given these relations, it can be shown that, apart from the already known representations that are inherited from $H(4)$ \cite{Baake:1982ah}, exactly five further inequivalent irreducible representations exist: $\tau^{\underline{4}}_1$, $\tau^{\underline{4}}_2$, $\tau^{\underline{8}}$, $\tau^{\underline{12}}_1$ and $\tau^{\underline{12}}_2$. The superscript denotes their dimension and the subscript numbers inequivalent irreducible representations of the same dimension.

While the representation $\tau^{\underline{4}}_1$ describes how spinors and thus quark fields are transforming under group action, Lorentz vectors such as derivatives transform according to the representation $\tau^4_1$ inherited from $H(4)$. Thus one might in principle construct the transformation matrices for three-quark operators under action of any given group element $G$:
   \begin{equation}
       \mathcal{O}^{(j),\text{transformed}} = G_{ij} \mathcal{O}^{(i)}.
   \label{groupaction}
   \end{equation}
However, this would again yield matrices of quite unhandy dimension (remember the amount of independent operators involved), so in a first step we reduced the set of possibly mixing operators and thereby the dimension of the transformation matrices by a detour via $SO_4$. Writing the quark fields with dotted and undotted indices in the chiral Weyl representation \cite{Peskin:1979mn} we can construct irreducible representations due to the local homomorphism $SU(2) \times SU(2) \simeq SO(4)$. Each quark spinor naturally consists of two two-component spinors, and after contraction with Pauli matrices also the covariant derivatives can be written in an $SU(2)$ representation: $\mathcal{D}_\mu \to {(\mathcal{D}_\mu \sigma^{\mu})^a}_{\dot b}$. For any given operator, independent symmetrization of dotted and undotted indices projects onto an irreducibly transforming leading twist operator, e.g.:
\begin{equation}
     u_{\dot a} u^{b} \mathcal{D}_\mu d^{c} \, \to \, u_{\dot a} u^{b} {(\mathcal{D}\sigma)^d}_{\dot e} d^{c} \, \to \,  u_{\{\dot a} u^{\{b} {(\mathcal{D}\sigma)^d}_{\dot e\}} d^{c\}}.
\end{equation}
Fixing the indices in the above example one can read off twelve independent operators belonging to one $SO_4$-irreducible representation. The chirality partners with exchanged dotted and undotted indices on the quark fields are treated in the same way.

It is now straightforward to construct irreducible representations for the symmetry group $O_4$ of the Euclidean continuum. As $O_4 = SO_4 \cup r SO_4$ with $r$ representing some reflection operation \cite{Baake:1982ah}, joining the parity partners of the $SO_4$ irreducible multiplets into one generates an $O_4$ irreducibly transforming multiplet.

At that point the number of possibly mixing operators is sufficiently reduced to construct their 768 transformation matrices $G_{ij}$ in eq.~(\ref{groupaction}). Forming suitable linear combinations with the help of the characters of the spinorial hypercubic group $\overline{H}(4)$ we construct a projector $P^\alpha$ that, applied to each $O_4$ irreducible operator multiplet, projects out such operators that are in fact $\overline{H}(4)$-irreducible:
  \begin{equation}
   P^\alpha=\frac{d_\alpha}{\vert \overline{H}(4) \vert}\sum_{G \in \overline{H}(4)} {\chi^\alpha(G)}^* \cdot G.
  \end{equation}
Here $d_\alpha$ denotes the dimension of the $\overline{H}(4)$-irreducible representation $\tau^\alpha$ to be projected on, $\chi^\alpha(G)$ the character of the group element $G$ in that representation and $\vert \overline{H}(4) \vert = 768$ is the group order.

We present two typical examples for $\overline{H}(4)$ irreducible operators. First we display a subset of a $\tau^{\underline{12}}_1$ irreducible multiplet of operators with one derivative:
 \begin{align}
  \mathcal{O}^{-+D+}_{1} &= -\frac{\sqrt{3}}{2\sqrt{2}} u_{\{\dot 0} u^{\{1} {(\mathcal{D}\sigma)^0}_{\dot 0\}} d^{0\}}  -\frac {\sqrt{3}}{2\sqrt{2}} u_{\{\dot 1} u^{\{1} {(\mathcal{D}\sigma)^1}_{\dot 1\}} d^{1\}}, \nonumber\\
  \mathcal{O}^{-+D+}_{2} &= \sqrt{3} u_{\{\dot 1} u^{\{1} {(\mathcal{D}\sigma)^0}_{\dot 0\}} d^{0\}}, \nonumber\\
  \mathcal{O}^{-+D+}_{3} &= -\frac{\sqrt{3}}{2\sqrt{2}} u_{\{\dot 0} u^{\{1} {(\mathcal{D}\sigma)^1}_{\dot 0\}} d^{1\}} -\frac{\sqrt{3}}{2\sqrt{2}} u_{\{\dot 1} u^{\{1} {(\mathcal{D}\sigma)^0}_{\dot 1\}} d^{0\}}, \nonumber\\
    \vdots \nonumber\\
  \mathcal{O}^{-+D+}_{12} &= -\frac{\sqrt{3}}{2\sqrt{2}} u^{\{0} u_{\{\dot 0} {(\mathcal{D}\sigma)^{0\}}}_{\dot 0} d_{\dot 0\}} - \frac{\sqrt{3}}{2\sqrt{2}} u^{\{1} u_{\{\dot 1} {(\mathcal{D}\sigma)^{1\}}}_{\dot 1} d_{\dot 0\}}. \nonumber
 \end{align}
The curly brackets denote total symmetrization in the (un)dotted indices. As a second example we show operators with two derivatives from one of the $\tau^{\underline{12}}_2$ representations:
 \begin{align}
  \mathcal{O}^{++DD+}_{1} =& \frac{5i}{4\sqrt{6}} u^{\{ 1} u^{0} {(\mathcal{D}\sigma)^0}_{\{ \dot 0} {(\mathcal{D}\sigma)^0}_{\dot 0 \}} d^{0 \}} - \frac{i\sqrt{3}}{4\sqrt{2}} u^{\{ 1} u^{1} {(\mathcal{D}\sigma)^1}_{\{ \dot 0} {(\mathcal{D}\sigma)^1}_{\dot 0 \}} d^{1 \}} \nonumber\\ 
                           & + \frac{5i}{2\sqrt{6}} u^{\{ 1} u^{1} {(\mathcal{D}\sigma)^1}_{\{ \dot 0} {(\mathcal{D}\sigma)^1}_{\dot 0 \}} d^{1 \}}, \nonumber\\
  \vdots \nonumber\\
  \mathcal{O}^{++DD+}_{12} =& -\frac{i}{2\sqrt{3}} u_{\{ \dot 0} u_{\dot 0} {(\mathcal{D}\sigma)^{\{1}}_{\dot 0} {(\mathcal{D}\sigma)^{0 \}}}_{\dot 0} d_{\dot 0 \}} - \frac{5i}{2\sqrt{3}} u_{\{ \dot 1} u_{\dot 1} {(\mathcal{D}\sigma)^{\{1}}_{\dot 1} {(\mathcal{D}\sigma)^{0 \}}}_{\dot 1} d_{\dot 0 \}}. \nonumber
 \end{align}

\begin{table}
 \renewcommand{\arraystretch}{1.4}
 \centerline
{
 \begin{tabular}{|c||c|c|c|}
  \hline
  & (mass)dimension 9/2 & dimension 11/2 & dimension 13/2\\
  \hline
  & 0 derivatives & 1 derivative & 2 derivatives \\
  \hline
  \hline
  $\tau_1^{\underline{4}}$ & 5 multiplets  & & 3 multiplets\\
  \hline
  $\tau_2^{\underline{4}}$ &  &  & 3 multiplets \\
  \hline
  $\tau^{\underline{8}}$ & 1 multiplet &  1 multiplet & 3 multiplets \\
  \hline
  $\tau_1^{\underline{12}}$ & 3 multiplets & 3 multiplets & 4 multiplets \\
  \hline
  $\tau_2^{\underline{12}}$ &  &  4 multiplets & 5 multiplets \\
  \hline
 \end{tabular}
}
\caption{Number of equivalent $\overline{H}(4)$ irreducible multiplets for three-quark operators with zero to two derivatives (not isospin symmetrized, derivatives acting on the last quark only).
   \label{tbl_irepr}}
\end{table}
Ordering the multiplets of irreducibly transforming three-quark operators according to their mass-dimension (number of derivatives) yields table~\ref{tbl_irepr}. This facilitates reading off the important results because mixing is possible only within one representation. As already mentioned, on the lattice operators of the same representation may additionally mix with those of lower mass dimension, e.g. involving an extra factor of $1/a$:
   \begin{equation}
      \mathcal{O}^{(i),\text{ren}}= Z_{ij} \mathcal{O}^{(j),\text{bare}} + Z' \cdot \frac{1}{a} \cdot  \mathcal{O}^{\text{bare, lower dim}}
   \end{equation}
However it is difficult and numerically challenging to factor out the divergent part. Therefore such operators should be avoided wherever possible or otherwise studied in detail so that the divergence can be cleanly isolated and finally subtracted from the renormalized operator.

In our approach to the nucleon distribution amplitudes we exclusively use the representations $\tau^{\underline{12}}_1$ for three-quark operators without derivative, $\tau^{\underline{12}}_2$ and $\tau^{\underline{4}}_2$ for three-quark operators with one and two derivatives, respectively. Thus we safely circumvent any power divergences in terms of the lattice spacing.

\section{Irreducible representations of the nucleon distribution amplitude operators}

In the last  section three-quark operators which are suitable for the calculation of matrix elements on the lattice were derived. The next task is to establish a connection between these three-quark operators and the quark distribution operators we are actually interested in. 
Before we give the corresponding relations, let us consider the additional symmetries among the moments of the quark distribution amplitudes introduced by the presence of two $u$-quarks in the nucleon:
   \begin{equation}
   V^{lmn}=V^{mln},\quad A^{lmn}=-A^{mln},\quad T^{lmn}=T^{mln}.
   \end{equation}
If we define
   \begin{equation}
   \phi^{lmn}=V^{lmn}-A^{lmn}+2T^{lnm},
   \end{equation} 
which is a natural combination in our analysis, then the fact that the nucleon has isospin $1/2$ implies
   \begin{equation}
   T^{lmn}=\frac{1}{6}(\phi^{lnm}+\phi^{mnl}).
   \end{equation}
With the analogous identities for $V$ and $A$ we can express the moments of $V$, $A$ and $T$ in terms of only one independent distribution amplitude $\phi^{lmn}$:
   \begin{align}
          V^{lmn}=&\frac{1}{6} \left(2 \phi ^{lmn}+2 \phi ^{mln}-\phi ^{nlm}-\phi^{nml}\right),\\ 
          A^{lmn}=&\frac{1}{6} \left(-2 \phi ^{lmn}+2 \phi ^{mln}-\phi^{nlm}+\phi^{nml}\right).
   \end{align}
The combination\footnote{Note that the normalisation conditions of $\varphi^{000}=1$ and $\phi^{000}=3$ are different} $\varphi^{lmn}=V^{lmn}-A^{lmn}$, often used in QCD sum rule calculations, can easily be expressed in terms of $\phi^{lmn}$:
   \begin{equation}
   \varphi^{lmn}=\frac{1}{3}\left(2\phi^{lmn}-\phi^{nml}\right).
   \end{equation} 
Due to momentum conservation there exists also a connection between lower and higher moments,
\begin{equation}
   \phi^{lmn}=\phi^{(l+1)mn}+\phi^{l(m+1)n}+\phi^{lm(n+1)}, \label{sumrule}
\end{equation}
allowing us to test our calculation.

Finally we relate the irreducible three-quark operators to the operators for the moments of the distribution amplitudes obtaining, e.g.,
   \begin{equation}
   \begin{split}
      \mathcal O^{+D-+}_{41}=&\frac{1}{2}\epsilon^{abc}
      \left(
            - u^a_1 (\mathcal{D}_3 u^b_3) d^c_1+
            i u^a_1 (\mathcal{D}_4 u^b_3) d^c_1+
              u^a_1 (\mathcal{D}_1 u^b_4) d^c_1-
            i u^a_1 (\mathcal{D}_2 u^b_4) d^c_1 
      \right)
      \\ =&
      \frac{1}{8}  
         \left(
               -\mathcal{A}_1^{13}+i \mathcal{A}_1^{14}+i \mathcal{A}_1^{23}+\mathcal{A}_1^{24}-\mathcal{A}_1^{31}+
               i\mathcal{A}_1^{32}+i \mathcal{A}_1^{41}+  \mathcal{A}_1^{42}
         \right.\\
         &\qquad
         \left.
               +\mathcal{V}_1^{13}-i \mathcal{V}_1^{14}-i \mathcal{V}_1^{23}-\mathcal{V}_1^{24}+\mathcal{V}_1^{31}-
               i\mathcal{V}_1^{32}-i \mathcal{V}_1^{41}-\mathcal{V}_1^{42}
         \right),
   \end{split}
   \end{equation}
where the lower index in the nucleon distribution amplitude operators denotes the spinor index and upper indices are Lorentz indices (see eq.~(\ref{eq_dav})-(\ref{eq_dat})). The symmetry in the Lorentz indices on the right-hand side reflects the leading twist projection of the operator. It is an element of the $\tau^{\underline{12}}_2$ representation and hence does not mix with any operators of lower dimension.
To construct an isospin 1/2 operator we use an operator from another, equivalent $\tau^{\underline{12}}_2$ representation,
   \begin{equation}
      \begin{split}
      \mathcal O^{++D-}_{65}=&\frac{1}{2}\epsilon^{abc}
      \left(
            - u^a_1 u^b_1 (\mathcal{D}_3 d^c_3)+i u^a_1 u^b_1 (\mathcal{D}_4 d^c_3) + 
              u^a_1 u^b_1 (\mathcal{D}_1 d^c_4)-i u^a_1 u^b_1 (\mathcal{D}_2 d^c_4) 
      \right)\\
      =&\frac{1}{8}  \left(\mathcal{T}_1^{13}-i \mathcal{T}_1^{14}-i \
\mathcal{T}_1^{23}-\mathcal{T}_1^{24}+\mathcal{T}_1^{31}-i \
\mathcal{T}_1^{32}-i \mathcal{T}_1^{41}-\mathcal{T}_1^{42}\right).
      \end{split}
   \end{equation}
Combining both operators we obtain an operator of isospin 1/2 with its matrix element given by
\begin{equation}
-4\langle 0|\mathcal{O}^{+D-+}_{41}-\mathcal{O}^{++D-}_{65}|p \rangle =
N_1 \left(p_1-i p_2\right) \left(p_3-i p_4\right) \mathit{f}_N \left(V^{\text{010}}-A^{\text{010}}+2 T^{\text{001}}\right).
\end{equation}

In the same way we deduced the following set of operators used in our analysis of the nucleon distribution amplitudes moments on the lattice 
\begin{itemize}
\item $0$ derivatives ($\tau^{\underline{12}}_1$) 
      \begin{align}
         \langle 0|\mathcal O^{12}_0 | N(p)\rangle&= f_N 
         (p_1\gamma_1-p_2\gamma_2)N(p),\quad  \nonumber\\
         \langle 0|\mathcal O^{34}_0 | N(p)\rangle&= f_N 
         (p_3\gamma_3-p_4\gamma_4)N(p),\nonumber\\
         \langle 0|\mathcal O^{1234}_0 | N(p),\rangle&=f_N 
         (p_1\gamma_1+p_2\gamma_2-p_3\gamma_3-p_4\gamma_4)N(p),\nonumber
      \end{align}
\item 
$1$ derivative ($\tau^{\underline{12}}_2$)
      \begin{align}
         \langle 0|\mathcal O^{12}_1 | N(p)\rangle=&f_N{\phi^{100}} 
            \left[
                  (\gamma_1 p_1-\gamma_2 p_2)(\gamma_3 p_3+\gamma_4 p_4)-2p_1p_2\gamma_1\gamma_2
            \right] N(p), \nonumber\\
         \langle 0|\mathcal O^{34}_1 | N(p)\rangle=&f_N{\phi^{100}} 
             \left[
                  (\gamma_1 p_1+\gamma_2 p_2)(\gamma_3 p_3-\gamma_4 p_4)-2p_3p_4\gamma_3\gamma_4
            \right] N(p), \nonumber\\
         \langle 0|\mathcal O^{1234}_1 | N(p)\rangle=&f_N{\phi^{100}}
         (\gamma_1 p_1-\gamma_2 p_2)(\gamma_3 p_3-\gamma_4 p_4)
         N(p), \nonumber
      \end{align} 
\item $2$ derivatives ($\tau^{\underline{4}}_2$)
\begin{align}
         \langle 0|\mathcal O^{1234}_2|N(p)\rangle&= f_N  {\phi^{110}} 
            \left[
                  p_1p_2\gamma_1\gamma_2\left(p_3\gamma_3-p_4\gamma_4\right)+
                  p_3p_4\gamma_3\gamma_4\left(p_1\gamma_1-p_2\gamma_2\right)
            \right]  N(p)\nonumber
      \end{align} 
\end{itemize}
with analogous expressions for other moments. The operators $\mathcal O_k$ denote  isospin $1/2$ combinations with $k$ derivatives carrying an implicit spinor index. For the $0$th moment we use only those operators which do not require non-zero spatial momenta. In the case of higher moments the advantage of using irreducible representations is paid by the requirement of non-zero spatial momenta. In the case of one derivative we used $p_1\neq0$ and for two derivatives $p_2,p_3\neq0$. It is important to notice that for the $0$th and $1$st moments we can use several operators thus improving our statistics. Unfortunately for the $2$nd moment we have to use operators from the $\tau^{\underline 4}_2$ representation to avoid mixing with lower-dimensional operators and hence we cannot increase our statistics in the same way.
\section{Unrenormalized results for the nucleon distribution amplitudes}
\begin{figure}
\centering
\includegraphics[clip,width=0.62\textwidth]{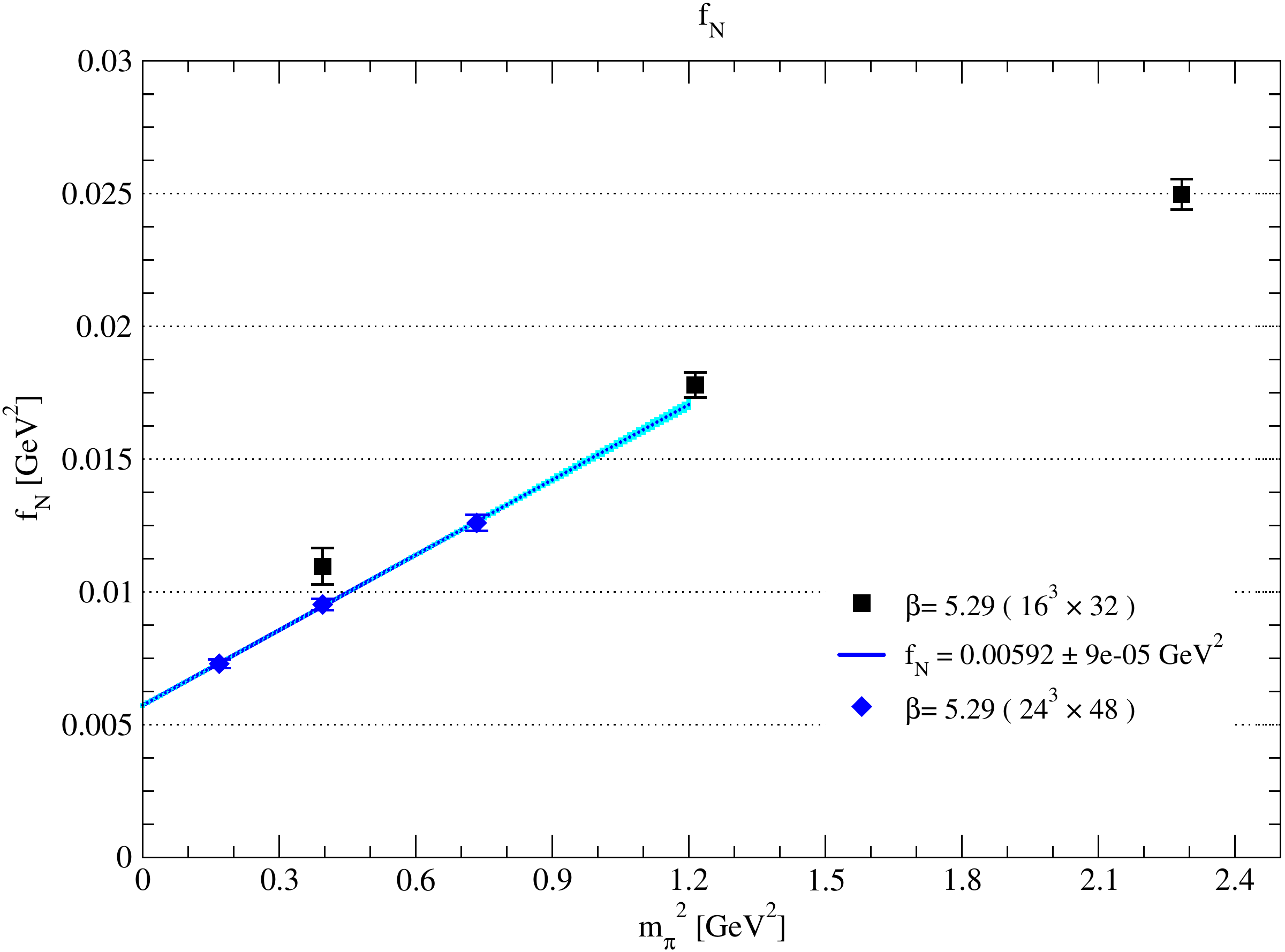}\\
\includegraphics[clip,width=0.62\textwidth]{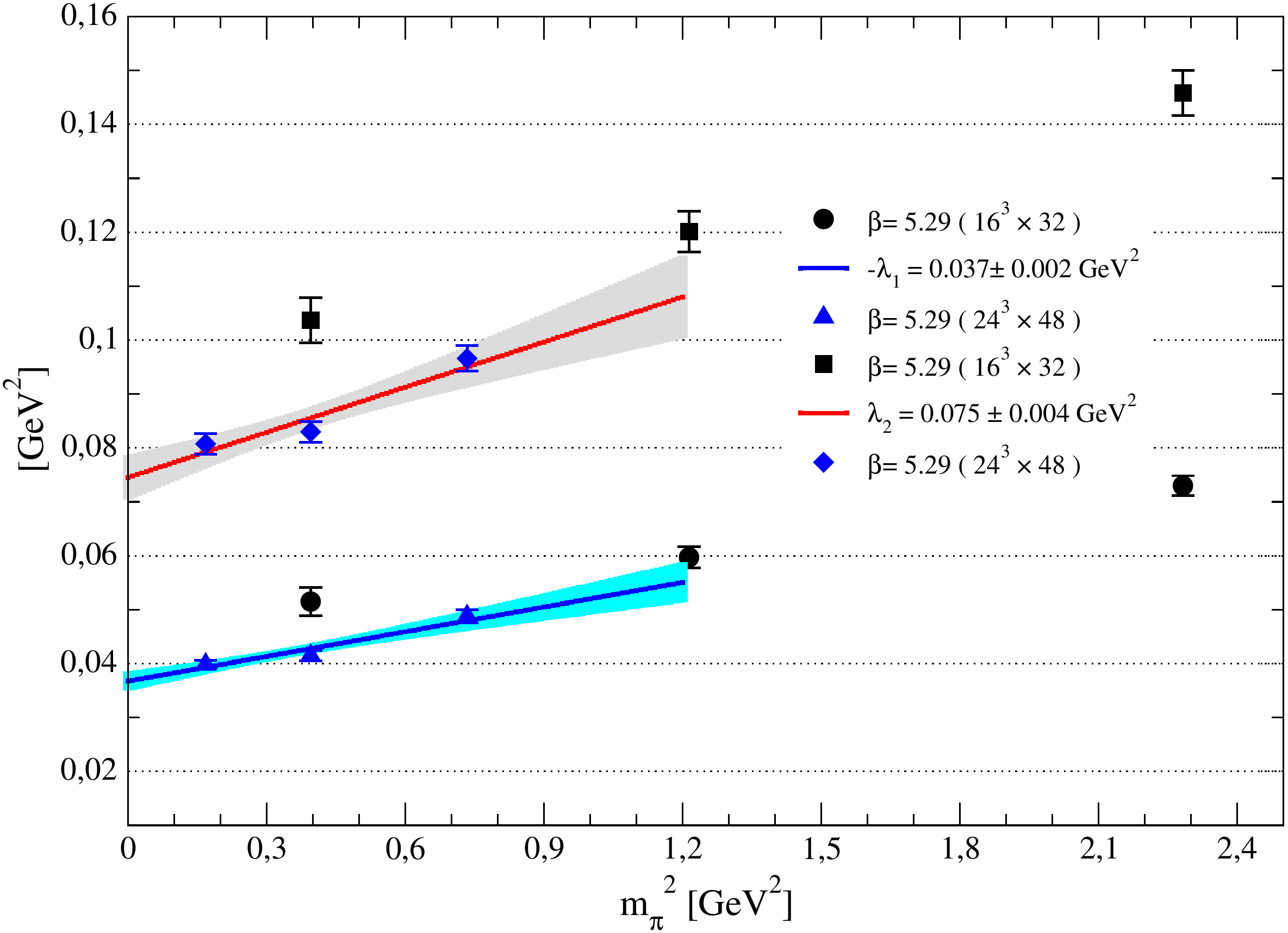}
\caption{\label{fig_mom0}
Nucleon decay constant $f_N$ (upper plot) and the normalization constants $-\lambda_1$ and $\lambda_2$ of the next-to-leading twist distribution amplidtudes (lower plot). All displayed results are unrenormalized.
}
\end{figure}
\begin{figure}
\centering
\includegraphics[clip,width=0.62\textwidth]{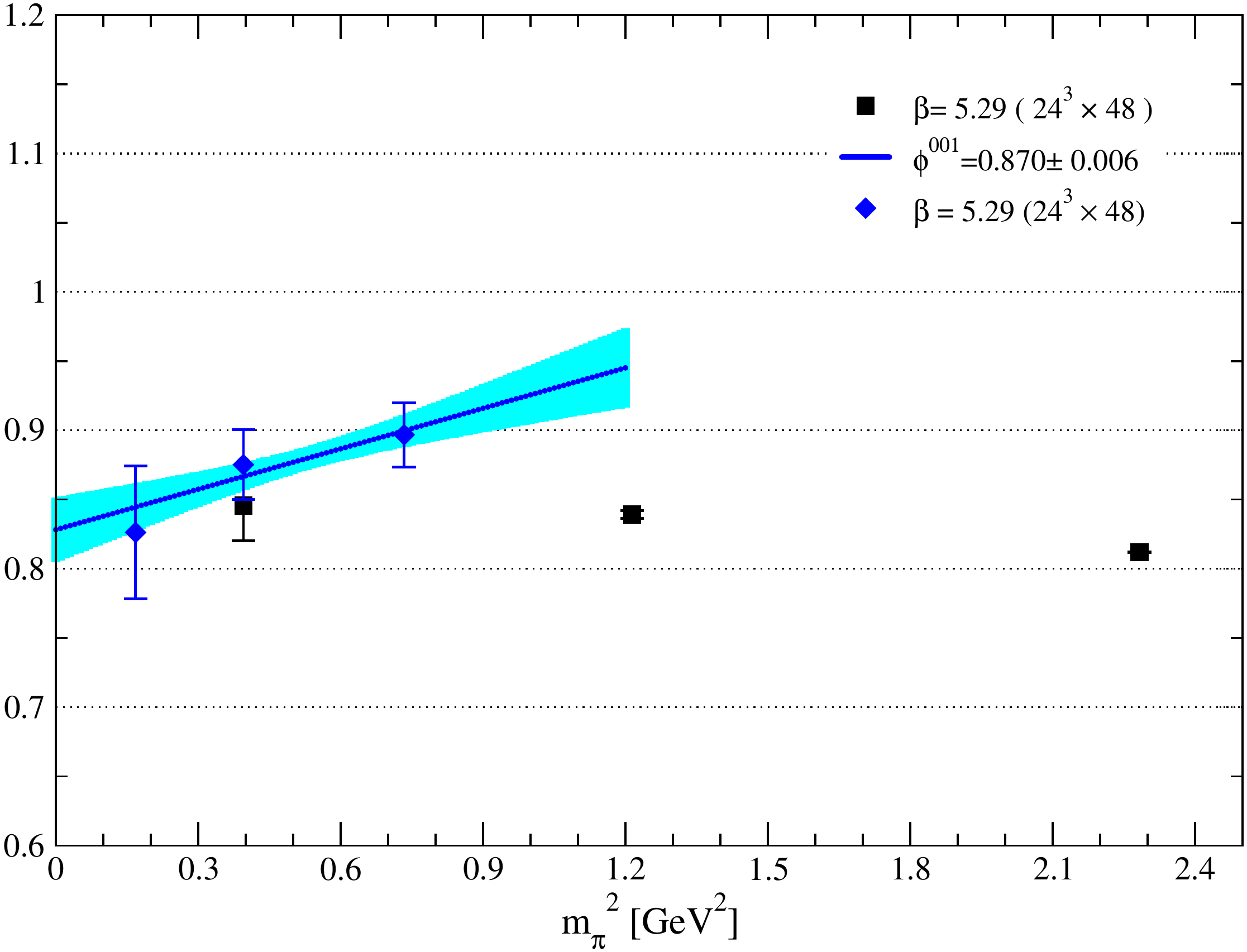}\\
\includegraphics[clip,width=0.62\textwidth]{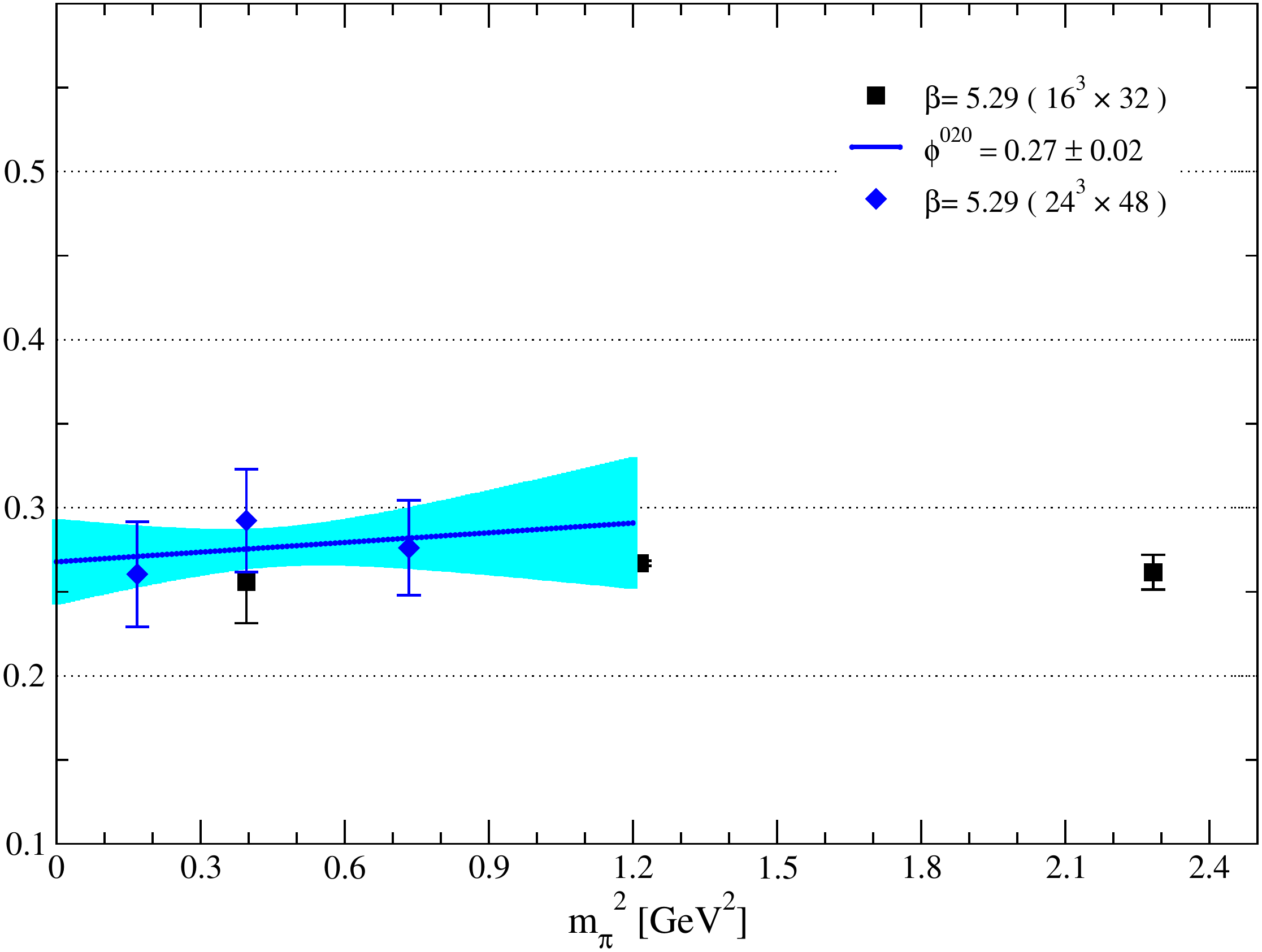}
\caption{\label{fig_mom12}
$\phi^{001}$ (upper plot) and $\phi^{020}$ (lower plot) moments of the leading twist nucleon dsitribution amplitude. All displayed results are unrenormalized.
}
\end{figure}

After these preparatory works we are finally able to obtain the desired moments of nucleon distribution amplitudes. We performed a lattice calculating of the above matrix elements, using parity and momentum projection for the appropriately chosen spatial momentum combinations. For the interpolating operator of the nucleon field we chose $\mathcal N_\tau=\epsilon^{abc} (u^aC\gamma_5 d^b) u^c_\tau$ leading to e.g. 
   \begin{equation}
      \langle \gamma_1 \gamma_4 \mathcal O^{12}_1(t)   \bar{\mathcal N} (0,\vec p) \rangle=f_N{\phi^{100}} \sqrt{Z_{N}(\vec p)}
       \frac{p_1 (E(\vec p) (m_N+E(\vec p))-2 p_2^2+p_3^2)}{E(\vec p)} \exp{\left[-E(\vec p) t\right]}
   \end{equation}
where parity projection is implied. 
The normalization constant $Z_{N}(p)$ can be extracted from the usual nucleon 2-point correlation function 
\begin{equation}
\langle \mathcal N(t,\vec p)   \bar{\mathcal N} (0,\vec p) \rangle=
Z_{N}(\vec p)\frac{m_N+E(\vec p)}{E(\vec p)} \exp{\left[-E(\vec p) t\right]}.
\end{equation} 

Since our operators for the moments of the nucleon distribution amplitudes are local we apply smearing only to the nucleon interpolating operator at the source. The two-point nucleon correlator is smeared both at source and sink. Due to this procedure the location of the effective mass plateaus is different for the two-point nucleon correlator and the two-point correlators used to determine the moments. Hence we do not calculate the ratios of these operators, but perform fully correlated fits to the correlators themselves choosing the fit ranges to match the effective mass plateaus. We use the pion masses determined by the QCDSF collaboration and set the scale with the Sommer parameter $r_0=0.467\textrm{fm}$. 

For the sake of flexibility we perform a two stage analysis. In the first step we calculate two-point functions of general 3-quark operators schematically given by
   \begin{equation}
      \langle 0|\epsilon^{abc}[\mathcal D^{\lambda_1}\dots \mathcal D^{\lambda_l} u]^a_\alpha 
                              [\mathcal D^{\mu_1}\dots \mathcal D^{\mu_m} u]^b_\beta 
                              [\mathcal D^{\nu_1}\dots \mathcal D^{\nu_n} d]^c_\gamma  \bar{\mathcal N}_\tau (p)|0\rangle,
   \end{equation}
with $l+m+n\leq2$. In a second step, these general operators can be used to calculate different matrix elements, in our case the irreducible combinations relevant for the moments of nucleon distribution amplitudes.

Although not discussed here, the calculation of moments of higher twist distribution amplitudes follows the same procedure as for leading twist. Thus using the general three-quark operators from step one, we can extract, e.g., the normalization constants $\lambda_1$ and $\lambda_2$ of the next-to-leading twist distribution amplitudes, which are also related to matrix elements of local three-quark operators \cite{Braun:2000kw}. 

Some of our results along with simple linear chiral extrapolations are presented in Figs.~\ref{fig_mom0}~and~\ref{fig_mom12}. In Fig.~\ref{fig_mom0} (upper plot) we show the normalization constant of the nucleon distribution amplitude  as function of the pion mass. The analysis was done on two different volumes, $16^3\times 32$ and $24^3\times 48$. We observe small finite size effects on the $16^3 \times 32$ lattice for the smallest pion mass. In the lower plot of Fig.~\ref{fig_mom0} we give $-\lambda_1$ and $\lambda_2$, the normalization constants of the next-to-leading twist distribution amplitudes. Compared to the leading twist case, we observe more pronounced finite size effects for these quanitites.

In Fig.~\ref{fig_mom12} we give representative results for two of the higher moments, $\phi^{001}$ (upper plot) and $\phi^{020}$ (lower plot), as a function of the pion mass. The data for the $16^3\times 32$ and $24^2\times 48$ lattices do not seem to be completely consistent. Improving statistics, we are currently examining this behavior, which is observed only for some of the higher moments. 

The results obtained on the  $24^3\times 48$ lattices are extrapolated linearly to the chiral limit. The resulting values for the normalization constants $f_N$, $\lambda_1$ and $\lambda_2$ are close to the QCD  sum rules calculations \cite{Chernyak:1984bm, King:1986wi}, but after renormalization we expect them to be lowered by approximately twenty percent.

\section{Renormalization}

All results presented so far refer to unrenormalized lattice calculations. The important step of renormalization for the three-quark operators used is in progress and will be discussed in a forthcoming paper, while here we will only sketch the keystones. 
For the renormalization we adopt a nonperturbative procedure, analogous to the well-established $RI-MOM$ scheme Martinelli et al.\ introduced for quark-antiquark operators \cite{Martinelli:1994ty}. We contract our isospin symmetrized and color antisymmetrized three-quark operators with three quark momentum sources on the lattice and calculate a correlation function of the following kind:
   \begin{equation}
     G^{(i)}(p,q,r)_{\alpha \beta \gamma}^{abc} =
      \int \!\! du \, dv \, dw \, dx \, e^{i (r\cdot u + p\cdot v + q\cdot w)} e^{-i(p+q+r)\cdot x} \langle \bar u(u)_{\alpha}^a \bar u(v)_{\beta}^b  \bar d(w)_{\gamma}^c \cdot \mathcal{O}^{(i)}(x) \rangle,
   \end{equation}
with $\alpha$, $\beta$ and $\gamma$ denoting spinor indices and $p$, $q$ and $r$ being the incoming quark momenta. The gauge is fixed to Landau gauge. After amputating the external legs, we are left with a three-quark vertex $\Gamma$ that contains all radiative corrections:
    \begin{equation}
     G^{(i)}(p,q,r)_{\alpha \beta \gamma}^{abc} =
     \Gamma^{(i)}(p,q,r)_{\alpha' \beta' \gamma'} S(-r)_{\alpha' \alpha} S(-p)_{\beta' \beta} S(-q)_{\gamma' \gamma} \epsilon^{abc}.
    \end{equation}
Imposing a suitable renormalization condition at the scale $\mu^2= (p^2+q^2+r^2)/3$ we obtain the renormalized vertex
        \begin{equation} 
          \Gamma^{(i),\text{ren}}(p,q,r;\mu) = Z^{\mathcal{O}}_{ij}(\mu) \cdot Z_q^{-3/2}(\mu) \cdot \Gamma^{(j)}(p,q,r).
        \end{equation}
The $RI-MOM$ renormalization matrix $Z^{\mathcal{O}}_{ij}$ is then converted to the $\overline{\mathrm{MS}}$ scheme using one loop continuum perturbation theory. Finally we can perform a renormalization group extrapolation to any renormalization scale $\mu$ we like.

\begin{figure}
   \centerline{
   \includegraphics[width=0.60\textwidth,clip]{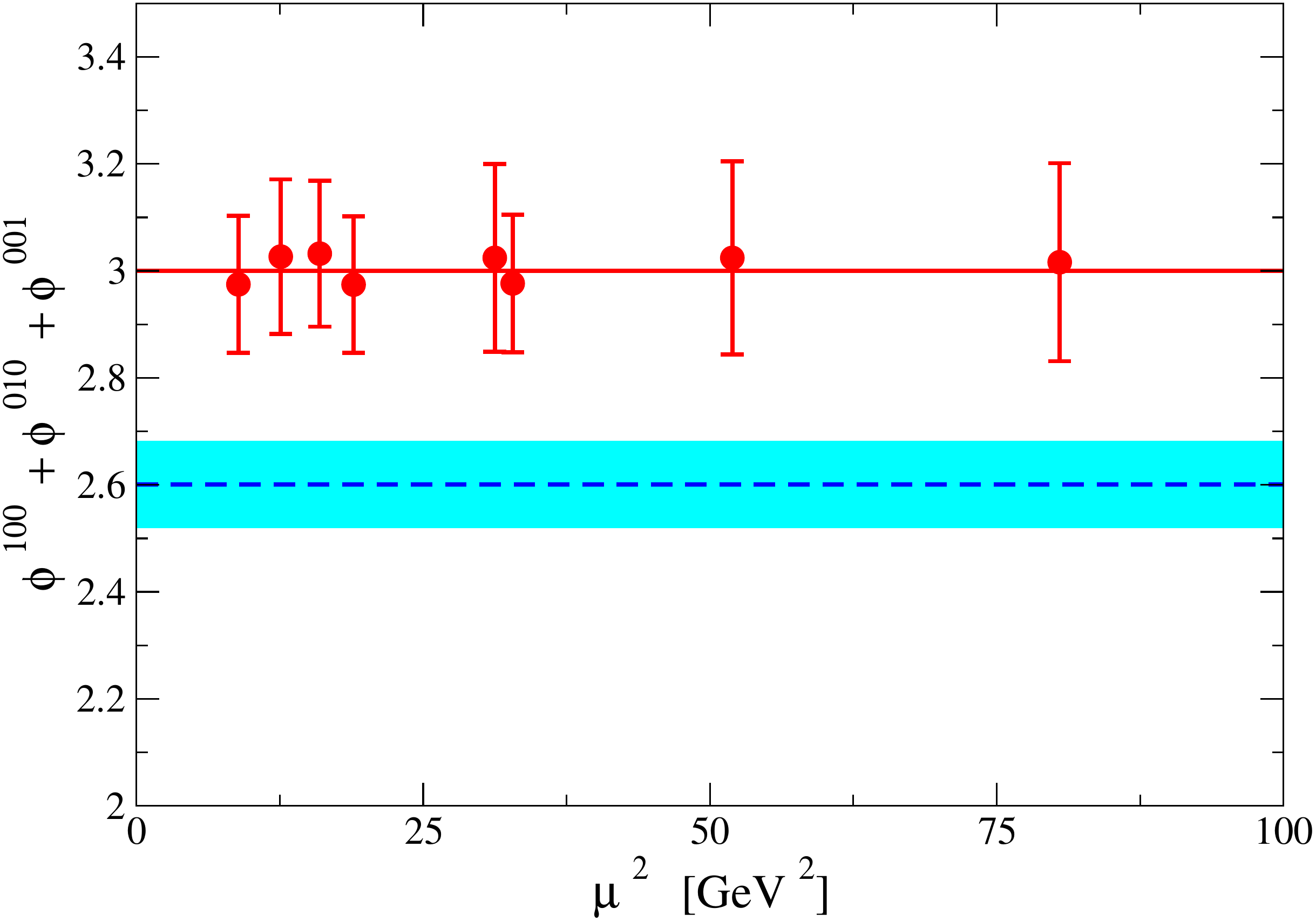}
   }
   \caption{Consistency check for the zeroth and first moments. Dashed line: sum of the bare first moments. Full line: renormalized sum of the first moments. The moments renormalized at different scales fit the theoretical constraint very well.
   \label{figsumrule}}
   \end{figure}

As an outlook and in order to demonstrate  how well this procedure works, we want to present a consistency check for the zeroth and first moments of our nucleon distribution amplitudes. Due to momentum conservation the sum of the three first moments is equal to the zeroth moment (cf. (\ref{sumrule})):
   \begin{equation}
      \phi^{100}+\phi^{010}+\phi^{001} = \phi^{000} = 3.
   \end{equation}
One can add up the bare first lattice moments and will receive a value of $2.60$, which is incompatible with the above sum rule (fig.~\ref{figsumrule}, dashed line). 
Incorporating the three-quark operator renormalization and especially the mixing matrix elements (we have to deal with a $3 \times 3$ matrix in this case) we obtain very good matching. We would like to emphasize that the well reproduced mean value of $3.00$ results from a renormalization at different scales and that the error bars containing uncertainties in the $\phi$s as well as in the $Z$s are a rather conservative estimate at this stage. This check makes us confident that our preliminary results tend into the right direction and deserve further investigations.

\section{Conclusions and perspectives}
In this paper we have discussed how moments of nucleon distribution amplitudes can be calculated on the lattice. After rewriting the lightcone expressions in terms of local matrix elements, we focused on operators that are well-behaved even in the context of reduced lattice symmetries. The representations of the spinorial hypercubic group provide the tool to derive irreducible sets of three-quark operators that are suited to control mixing and renormalization.
After appropriate isospin symmetrization we expressed the operators for the moments of the nucleon distribution amplitudes in terms of these irreducible representations and showed preliminary unrenormalized results for various moments of the distribution amplitudes. The renormalized results for the first moments are in good agreement with a sum rule.
A comprehensive and detailed study of the three-quark operator renormalization is in progress right now.

\acknowledgments
For our computations we employed QCDSF/UKQCD configurations. They have been
generated on the Hitachi SR8000 at LRZ (Munich), the Cray T3E at EPCC 
(Edinburgh) \cite{Allton:2001sk}, the APE{\it 1000} at NIC/DESY (Zeuthen) 
as well as the BlueGene/L at NIC/FZJ (J\"ulich). The renormalization 
matrices have been calculated on a QCDOC machine in Regensburg using 
USQCD software and Chroma \cite{Edwards:2004sx, bagel}. The general 
three-quark operators have been evaluated on the APE{\it 1000} at NIC/DESY 
(Zeuthen). This work has been supported in part by the BMBF, the DFG 
(Forschergruppe Gitter-Hadronen-Ph\"anomenologie) and
by the EU Integrated Infrastructure Initiative (I3HP) under contract
number RII3-CT-2004-506078.
We thankfully acknowledge helpful discussions with V.~Braun and 
A.~Lenz on QCD sum rules.

\end{document}